\documentclass[conference]{IEEEtran}

\usepackage{cite}
\usepackage{multirow}
\usepackage{tabularx}
\usepackage{makecell}
\usepackage{hanging}
\usepackage[]{graphicx}
\graphicspath{{figures/}}
\DeclareGraphicsExtensions{.eps,.pdf}

\usepackage{amsmath,amssymb,amsthm,dsfont}
\usepackage[mathscr]{euscript}
\usepackage{mathtools}
\usepackage{verbatim}
\newtheorem{theorem}{Theorem}

\newtheorem{lemma}{Lemma}

\theoremstyle{remark}
\newtheorem{remark}{Remark}

\allowdisplaybreaks[4] 
\usepackage{ctable}
\usepackage{color}

\IEEEoverridecommandlockouts

\renewcommand{\baselinestretch}{0.975}

\begin{document}

\title{Achievable Rate of Private Function Retrieval from MDS Coded Databases}

\author{ \thanks{This work is supported by NSF grant CNS-1526547.}
\IEEEauthorblockN{Sarah A. Obead, J\"{o}rg Kliewer}
\IEEEauthorblockA{Helen and John C.~Hartmann Department of Electrical and Computer Engineering\\
New Jersey Institute of Technology\\
Newark, New Jersey 07102\\
Email: sao23@njit.edu, jkliewer@njit.edu}
}

\maketitle

\begin{abstract}
We study the problem of private function retrieval (PFR) in a distributed storage system. In PFR the user wishes to retrieve a linear combination of $M$ messages stored in non-colluding $(N,K)$ MDS coded databases while revealing no information about the coefficients of the intended linear combination to any of the individual databases. We present an achievable scheme for MDS coded PFR with a rate that matches the capacity for coded private information retrieval derived recently, $R=(1+R_c+R_c^2+\dots+R_c^{M-1})^{-1}=\frac{1-R_c}{1-R_c^M}$, where $R_c=\frac{K}{N}$ is the rate of the MDS code. This achievable rate is tight in some special cases.
\end{abstract}

\vspace*{-1ex} 
\section{Introduction}
The private retrieval of information from public databases has received
significant attention already for several decades from researchers in the
computer science community (see, e.g., \cite{Chor_etal1998,
  Yekhanin2010}). While this line of work, commonly known as private information
retrieval (PIR), is concerned with downloading individual messages in a
private manner from databases, a recently proposed generalization of this
problem \cite{Maddah-Aliarxiv2017, SunJafararxiv_2017} addresses the private
computation of functions of these messages. In accordance with \cite{Maddah-Aliarxiv2017} we denote this approach as private
function retrieval (PFR) in the following.  In PFR a user has access
to a given number of databases and intends to compute a function of messages
stored in these databases. This function is kept private from the
databases, as they may be under the control of an eavesdropper. Both works \cite{Maddah-Aliarxiv2017, SunJafararxiv_2017}
characterize the fundamental information theoretic communication overhead
needed to reliably compute the given function and specify the corresponding
capacity and achievable rates as a function of the message size, the number
of messages, and the number of databases, respectively. Further, the authors
assume that the data is replicated on each database. Surprisingly, the obtained PFR capacity
result is equal to the PIR capacity of
\cite{Sun_Jafar2015}.

However, although repetition coding adds the largest amount of redundancy and
thus protects effectively against erasures, it is associated with a large
storage cost. A more general way to optimally trade-off the available
redundancy (or rate) versus the erasure correcting capability is given by MDS
codes. In particular, for an $(N,K)$ MDS code with $N$ code symbols and $K$
information symbols and rate $R_c=K/N$ $N-K$ erasures can be
recovered from any $K$ code symbols. Coded PIR has been
addressed in two different lines of work. Achievable schemes for MDS coded
PIR have been presented in \cite{Chanetal_ISIT2015, Tajeddineetal_ISIT2016}
and the capacity has been established in
\cite{BanawanUlukus_Globecom2017}. On the other hand, in \cite{Yaakoobi_ISIT2015} linear codes with $k$
different reconstruction sets for each code symbol have been proposed in
form  of so called $k$-server PIR.

In this paper we propose coded PFR, which to the best of our knowledge has
not been addressed yet in the recent literature, with the notable exception
of the parallel work in \cite{Karpuk}, which is based on a fixed ($k$-)
server PIR scheme with the inclusion of colluding databases. Our scheme is
based on MDS codes which in contrast to \cite{Karpuk} minimize the storage
overhead and maximize the achievable download rate. In particular, we provide a
characterization of the achievable rate of MDS coded PFR if the user wishes
to compute an arbitrary linear combination of $M$ independent equal-sized
messages over some finite field $\mathbb{F}_q$, distributed over $N$
non-colluding MDS-coded databases. Surprisingly, our achievable rate matches
the capacity for MDS coded PIR in \cite{BanawanUlukus_Globecom2017}. This
demonstrates that, compared to the naive scheme, where $M$ coded messages are
downloaded and linearly combined offline at the user (requiring $M$-times
the coded PIR rate), downloading the result of the computation privately and
directly from the databases does not incur any penalty in rate compared to
the coded PIR case. Thus, our result strictly generalizes the achievable schemes in
\cite{Maddah-Aliarxiv2017, SunJafararxiv_2017} which represent special
cases of our proposed PFR scheme.

\vspace*{-0.5ex}
\section{Problem Statement} \vspace*{-0.5ex}\label{ProblemStatment}
In the following, we use {$[1:X]$} to denote the set $\{1,\dots,X\}$. Similarly, $X_{1:N}=\{X_1,\dots, X_N\}$. 

\vspace{-0.5ex}
\subsection{System Model}\vspace{-0.5ex}
In coded PFR, a user wishes to privately retrieve a linear combination of the messages stored in the databases such that the coefficients of the linear combination are kept secret from each individual database. Consider a linear distributed storage system storing $M$ equal-sized messages on $N$ non-colluding  databases. The message $W_m, m\in [1:M],$ is composed from $L$ symbols chosen independently and uniformly at random from the finite field $\mathbb{F}_{q}$ with
\begin{align} 
H(W_1)=& \dots = H(W_M)=L \log q,\\ 
\hspace{-0.45ex}H(W_1,\!\dots\!,W_M)=& H(W_1)+\!\dots\! + H(W_M)=ML \log q. \vspace*{-0.5ex}
\end{align}

Each message is divided into $\tilde{L}$ segments, each of $K$ symbols, forming a $\tilde{L}\times K$ matrix, where $L=\tilde{L} K$. The  messages are stored using an $(N,K)$ MDS code with the full rank generator matrix defined by
\begin{equation} \vspace*{-0.5ex}
{\bf G}=\big[ {\bf{g}}_1\quad {\bf{g}}_2\quad \dots\quad {\bf{g}}_N\big]_{K\times N},
\end{equation}
with ${\bf{g}}_n,\;n\in[1:N],$ denoting the $n$-th column vector of ${\bf G}.$ 
The generator matrix produces a code that can tolerate up to $N-K$ erasures by retrieving data from any set ${\mathcal{K}}\subset \{1,\dots,N\}$ databases, where $|{\mathcal K}|\geq K$. The encoding process for message $W_m$ is defined as follows:
\begin{equation}
\hspace{-4ex}\begin{aligned}
{\begin{bmatrix}{\bf{w}}_{m,t}\end{bmatrix}}_{1\times K} &{\begin{bmatrix} {\bf{g}}_1& {\bf{g}}_2& \dots& {\bf{g}}_N\end{bmatrix}}_{K\times N}\\
&\quad\qquad= {\begin{bmatrix} {\bf{g}}_1^T{\bf{w}}_{m,t}&\dots&{\bf{g}}_N^T{\bf{w}}_{m,t}\\ \end{bmatrix}}_{1\times N}, 
\end{aligned}
\end{equation}
where ${\bf{w}}_{m,t},$ $\forall m\in[1:M], \forall t\in[1:\tilde{L}],$ denotes the $K$-dimensional vector of symbols of the $t$-th segment from the message $W_m$. The resulting $N$ coded symbols for each segment are then distributed over the $N$ databases, and the code rate is given by $R_c=\frac{K}{N}.$

Consequently, the code symbols stored at each database $n\in [1:N]$ are given by
\begin{equation}\label{eq:codedDatabse}
{\bf W}_{DB_n}=
\begin{bmatrix}
{\bf{g}}_n^T{\bf{w}}_{1,1} & {\bf{g}}_n^T{\bf{w}}_{1,2}  &\dots& {\bf{g}}_n^T{\bf{w}}_{1,\tilde{L}}\\
\vdots &\vdots&  \ddots & \vdots\\
{\bf{g}}_n^T{\bf{w}}_{M,1} & {\bf{g}}_n^T{\bf{w}}_{M,2}  &\dots& {\bf{g}}_n^T{\bf{w}}_{M,\tilde{L}}\\
\end{bmatrix},
\end{equation}
\noindent where we use ${\bf{W}}[t]$ to denote the $t$-th column, and $W_{m}(t)$ for the element of the $m$-th row  and $t$-th column of the database, respectively. 

In PFR, the linear combination $\nu$ the user intends to retrieve is represented as 
\begin{align}
\widetilde{W}_\nu &= {\bf{v_\nu} }[W_1,\dots,W_M]^T \label{eq:v_msg}\\
&=v_{\nu}(1)W_1+\dots+v_{\nu}(M)W_M \label{eq:v_msg1}\\ 
&= \begin{bmatrix}{\bf{v_\nu} }{\bf{W}}[1]& \dots & {\bf{v_\nu} }{\bf{W}}[\tilde{L}]\end{bmatrix}\!, \label{eq:v_ms3}
\end{align}
where ${\bf v_\nu}$ is an $M$-dimensional non-zero coefficient vector of the linear combination (row vector) indexed by $\nu$, the coefficients $v_{\nu}(m), \;\forall m\in[1:M],$ are chosen from the finite field $\mathbb{F}_{q}$, and the addition ``$+$''~is done element-wise over the same field. We assume that the vector ${\bf v_\nu}$ is an element of the set ${\cal V}$ that contains all possible distinct $M$-dimensional vectors defined over ${\mathbb F_q}$ where $\nu \in [1:V],\;V= |{\cal V}|=\frac{q^M-1}{q-1}$. 
 
 In order for the user to retrieve the linear combination $\widetilde{W}_\nu$, while keeping $\nu$ secret from each database, it generates $N$ query matrices for the databases $\{Q_1^{[\nu]},\dots,Q_N^{[\nu]}\}$. Since the query matrices are generated by the user without prior knowledge of the realizations of the stored messages, the queries must be independent of the messages,
 \begin{equation}\vspace*{-0.5ex}
 I(Q_1^{[\nu]},\dots,Q_N^{[\nu]};W_1,\dots,W_M)=0, \quad \forall \nu\in[1:V].
 \end{equation}
 
 Upon the reception of the query $Q_n^{[\nu]}$, the $n$-th database generates an answer string $A_n^{[\nu]}$ as a deterministic function of the received query and the stored symbols from each message. Hence, 
 \begin{equation}\vspace*{-0.5ex}
	\!H(A_n^{[\nu]}|Q_n^{[\nu]},{\bf W}_{DB_n})=0, \quad\! \forall \nu\in[1:V],\forall n\in[1:N]. \!\!
 \end{equation}
To maintain user privacy, the query-answer function must be identically distributed for each possible linear combination $\nu\in[1:V]$ from the perspective of each database $n\in[1:N]$. In other words, the scheme's queries and answers strings must be independent from the desired linear combination index, therefore the following privacy constraint must be satisfied:\vspace*{-1ex}
\begin{equation}\label{privacy_const}
\begin{aligned}
 I(A_n^{[\nu]},Q_n^{[\nu]},{\bf W}_{DB_n};\nu)=0, \!\quad\! \forall \nu\in[1:V].\!
\end{aligned}
\end{equation} 
After the user receives all answer strings from each database, the user must be able to reliably decode the desired linear combination message $\widetilde{W}_\nu$ with a probability of error $P_e$ that goes to zero as the message size $L$ approaches infinity. Following Fano's inequality, this translates to the decodability constraint
 \vspace*{-1ex}\begin{equation}\label{decoding_const}
 H(\widetilde{W}_\nu |A_{1:N}^{[\nu]},Q_{1:N}^{[\nu]})= o(L),
 \end{equation}
 
\noindent where $o(L)$ represents any function of $L$, $f(L)$,  that satisfies $\lim_{L\rightarrow \infty} f(L)/L\rightarrow 0.$
 
The retrieval rate of the coded PFR scheme is characterized by the message length $L$, the query structure $Q,$ and the query-answer function, and is defined as the ratio between the size of the desired linear combination message and the total number of downloaded symbols in bits as
\vspace*{-1ex}\begin{equation}\label{eq:PFR_rate_def} 
R=\frac{H(\widetilde{W}_\nu)}{\sum_{n=1}^{N} H(A_n^{[\nu]})} .
\end{equation}
 
A rate $R$ is said to be achievable if there exist a sequence of coded PFR schemes that satisfy the privacy and correctness constraints of \eqref{privacy_const}, \eqref{decoding_const} for $P_e\rightarrow 0$ as $L\rightarrow \infty$. 

\vspace*{-0.5ex}
\section{Achievable Rate of MDS Coded PFR} \label{MainResults}
\begin{theorem}
	For an $(N,K)$ coded distributed storage system with code rate $R_c=\frac{K}{N}$, $M$ messages and a set of $V$ linear combinations defined over the field $\mathbb{F}_{q}$, a PFR achievable rate is given as \vspace*{-2ex}
\begin{align}
\qquad \qquad R&\leq \frac{1-R_c}{1-R_c^M} \label{eq:capacity1}\\
&=\Big(1+\frac{K}{N}+\frac{K^2}{N^2}+\dots+\frac{K^{M-1}}{N^{M-1}}\Big)^{-1}. \label{eq:capacity}
\end{align}\vspace{-2ex}
\end{theorem}
\begin{remark} 
This achievable rate generalizes the achievable rate of repetition coded PFR \cite{SunJafararxiv_2017} which corresponds to the special case of $K\!=\!1$.
Also, \eqref{eq:capacity1} is only a function of the distributed storage
coding rate $R_c$ and the number of stored independent messages $M,$ and is
universal in the sense that it does not depend on the number of linear
combinations $V$ defined over the finite field $\mathbb{F}_{q}$ nor on the
explicit structure of the code.
\end{remark}
\begin{remark}
	If we consider each of the $V$ linear combinations of messages in \eqref{eq:v_msg} as a new \emph{virtual message} $\widetilde{W}_\nu$, and then apply the coded PIR scheme of \cite{BanawanUlukus_Globecom2017}, the scheme rate will be $\frac{1-R_c}{1-R_c^V}$ which is smaller than \eqref{eq:capacity1} since $M\leq V$. 
\end{remark}
\begin{remark}
  When the linear combination set ${\cal V}$ is reduced to the first $M$
  linear combinations
  (i.e.,~${\bf{v}}_{1:M} \in {\cal V}:
  [{\bf{v}}_1\;{\bf{v}}_2\;\dots\;{\bf{v}}_M]={\bf I}_M$),
  the achievable rate of \eqref{eq:capacity1} is tight. That is
  because in this setting the problem of coded PFR is reduced to coded PIR
  where the converse is implied from \cite{BanawanUlukus_Globecom2017}.
  Also, we note that \eqref{eq:capacity1} is equivalent to the coded PIR
  capacity \cite{BanawanUlukus_Globecom2017}, which has been observed in
  \cite{SunJafararxiv_2017} for $K=1$. Thus,
  downloading linear combinations of messages does not incur additional costs
  over downloading individual messages. 
\end{remark}
\begin{remark} 
	Eq. \eqref{eq:capacity1} is a strictly decreasing function in the number of messages $M$ for fixed $R_c$. As the number of messages increases $M\rightarrow \infty$, the achievable rate approaches $1-R_c$. Moreover, as $R_c \rightarrow 1$ in \eqref{eq:capacity}, $R \rightarrow \frac{1}{M},$ indicating that to maintain the privacy of the desired linear combination, the user must download all the messages and perform the computation off-line. 
\end{remark}

\vspace*{-1ex}
\section{Proof of Theorem~1}  \label{AchievableScheme}

\vspace*{-0.5ex}
\subsection{Query generation}
\vspace*{-0.5ex}
The generation of the queries is shown in Algorithm~1. 
Let $B\in[1:V]$ be the block indicator and $R\in[1:K]$ be the repetition indicator, respectively. Let the $v$-sum be the combination of $v$ distinct elements out of $V$ elements. Since we have $V \choose v$ different combinations, we denote each different combination as a \emph{type} of the $v$-sum. Let the components of these combinations be symbols of the $V$ virtual messages. As mentioned above, we generate the query set for each database in blocks, where a block represents a group of all  ${V\choose v}$ types of $v$-sums for all $v \in[1:V],$ resulting in $V$ blocks in total. To this end, we let the size of the \emph{dependent} virtual messages to be $L=KN^V$ (i.e.,~${\tilde{L}=N^V}$).

For a desired linear combination $\nu \in[1:V]$ we use the notation $Q^{[\nu]}(DB_B)$ to indicate the query set of the database $DB_B\in[1:N]$. This set is composed from $VK$ disjoint subsets $Q^{[\nu]}_{B,R}(DB_B)$ generated for each block $B$ and repetition $R$. We require $K^{V-B}(N-K)^{B-1}$ distinct instances of each type of $v$-sum for every set $Q^{[\nu]}_{B,R}(DB_B)$. Each block and repetition subset is further subdivided into two subsets: the first subset $Q^{[\nu]}_{B,R}(DB_B,{\cal M})$ consists of the $v$-sum types with symbols from the desired linear combination, and the second subset $Q^{[\nu]}_{B,R}(DB_B,{\cal I})$ contains only $v$-sum types with symbols from undesired linear combinations. The query sets for all databases are generated by Algorithm 1 with the following procedures. 

\indent {\bf \textit{1) Index assignment}}: In the MDS-coded PIR scheme
\cite{BanawanUlukus_Globecom2017}, the user privately applies a random
permutation over the coded symbols of each message independently. The goal
is to make the coded symbols queried from each database to appear to be
chosen randomly and independently from the desired message. However, for the
PFR problem the linear function is computed element-wise, thus there is a
dependency across the symbols with the same index, which must be maintained
under a permutation. To this end, we modify the permutation to be fixed
across all messages.  Let $\pi(\cdot)$ be a random permutation function over
$[1:\tilde{L}]$. We use the notation $U_\nu(t)$, where
	\begin{equation}
	U_\nu(t)\triangleq \sigma_t\widetilde{W}_\nu(\pi(t))=\sigma_t{\bf v}_{\nu}{\bf{W}}[\pi(t)],
	\end{equation} 
	to indicate the permuted message symbol from the virtual message $\widetilde{W}_\nu$. The random variable $\sigma$ is used to indicate the sign assigned to each individual virtual message symbol, $\sigma_t \in \{+1,-1\}$ \cite{SunJafararxiv_2017}. Both $\sigma_t$ and $\pi$ are randomly selected privately and uniformly by the user.

\indent {\bf \textit{2) Block ${B=1}$}}: This block is described by Steps {3} to {10} of Algorithm 1, where we have $v=1$ for the $v$-sum.\\	 
\indent {\textit{Initialization:}}
In the initialization step, the user queries the first database {${DB_1=1}$} for $K^{V-1}$ distinct symbols from the desired linear combination $U_\nu(i)$. This is done by calling the function "$\text{new}(U_\nu)$" that will select a symbol from message $U_\nu$ with a new index $i$ each time it is called (Step 6).

\indent {\textit{Database symmetry:}}
Database symmetry is obtained via the ``For'' loop in Step {3}, resulting in a total number of $N K^{V-1}$ symbols over all databases. 

\indent {\textit{Message symmetry:}}
In Step {7}, to maintain message symmetry, the user ask each database for the same number of distinct symbols of all other linear combinations $U_\theta(i),\;\theta\in\{1,\dots,V\}\!\setminus\!\{\nu\},$ resulting in a total number of $NVK^{V-1}$ symbols. As a result, the query sets for each database are symmetric with respect to all linear combination vectors in $[1:V]$. We associate the symbols of undesired messages in $K$ groups $G\in{[1:K]}$ to be exploited as distinct side information for different rounds of the scheme as shown in Step {7.}

\indent {\bf \textit{3) Side-information exploitation:}}
In Steps {11} to {20}, we generate the blocks {$B\in[2:V]$} by applying two subroutines ``Exploit-SI'' and ``M-Sym'', respectively. We first use the subroutine ``Exploit-SI'' \cite{SunJafararxiv_2017} to generate queries for new symbols of the desired linear combination $U_\nu$ by combining these symbols with different side information groups from the previous block associated with $N-K$ neighboring databases, as shown in Step {13}. This is required by our proposed MDS coded scheme to ensure privacy and is in contrast to \cite{SunJafararxiv_2017}, where the side information of previous blocks from \emph{all} databases is utilized. 

Then, the subroutine ``M-Sym'' \cite{SunJafararxiv_2017} is used to generate side information to be exploited in the following blocks. This subroutine select symbols of undesired messages to generate $v$-sums that enforce symmetry in the block queries. For example in $B=2$, if we have the queries $U_\nu(i)+U_2(j)$, and $U_\nu(l)+U_3(r) \in Q^{[\nu]}_{2,R}(DB_2,{\cal M})$, this subroutine will generate $U_2(l)+U_3(i)$. As a result, we can show that the symmetry over the linear combinations and databases is maintained. By the end of this step we have in total $N{V\choose B} K^{V-B}(N-K)^{B-1}$ queries for each block from all databases. 

\indent {\bf \textit{4) Generation of further query rounds:}} 
We require further query rounds to obtain $K$ linear equations for each coded symbol to be able to decode. To this end, we circularly shift the order of the database at each repetition. The shift is done for the initial block, $B=1$, in Steps {22} to {25}. However, for the following blocks we only rotate the indices of desired messages $U_\nu$ and combine them with new groups of side information from the neighboring databases from the first round as seen in Steps {26} to {33}. This rotation and side information exploitation for $B\in[2:V]$ is done using the subroutine ``Reuse-SI'' (omitted in the interest of space).

\indent {\bf \textit{5) Query set assembly:}} 
Finally, in Steps {35} to {37}, we assemble each query set from the queries disjoint subsets obtained in the previous blocks and rounds.

\begin{remark}
Note that the proposed scheme significantly differs from the one presented in \cite{SunJafararxiv_2017} in terms of how the side information is exploited due to coding. In particular, we distribute the side information over $K$ rounds such that every database is queried for each message and linear combination only once. 
\end{remark}  

\begin{table}[ht!]
	\small
	\begin{tabular}{p{0.965\linewidth}}
		\specialrule{.1em}{.05em}{.05em} 
		\rule{0pt}{2.5ex} 
		\hspace{-0.5em}\noindent\textbf{Algorithm 1:} Query set generation algorithm\\
		\specialrule{.1em}{.05em}{.05em} \specialrule{.1em}{.05em}{.05em} 
		\rule{0pt}{2.5ex}  
		\hspace{-0.35em}\textbf{Input:} $\nu, K, N, M,$ and $V.$\\	
		\textbf{Output:} ${Q}^{[\nu]}(1),\dots,{Q}^{[\nu]}(N)$\\
		\begin{hangparas}{.15in}{1}1.~\textbf{Initialize:} All query sets are initialized as a null set $Q^{[\nu]}(1),\dots,Q^{[\nu]}(N)\leftarrow \emptyset$, the block counter $B=1,$ and repetition counter $R=1.$ Let  number of neighboring databases $Nb= {N-K}$ \end{hangparas} 
		
		2.~\textbf{Let} repetition $R_{B}= K^{V-B}(N-K)^{B-1}\quad \forall B\in[1:V]$\\
		3.~\textbf{For} first database block $DB_1=1:N$ \textbf{do}\\ 
			4.\hspace{2ex}\textbf{For} side information group $G=1:K$ \textbf{do} \\
				5.\hspace{4ex}\textbf{For} repetition group $RG=1:(R_{1}/K)$ \textbf{do} \vspace{-1ex} 
					\begin{equation*}\vspace{-0.5ex}
						\begin{aligned}
				\hspace{-2ex}\text{6.} \qquad\qquad Q^{[\nu]}_{1,R}(DB_1,{\cal M})\!\leftarrow& \{u_{\nu}\}, u_\nu =\text{new}(U_{\nu})\\
				\hspace{-2.8ex}\text{7.} \qquad\qquad	Q^{[\nu]}_{1,R}(DB_1,{\cal I}_{G})\!\leftarrow& \{\text{new}(U_{1}),\dots,\text{new}(U_{V})\}\!\setminus\!\{u_{\nu}\}
						\end{aligned}
					\end{equation*}
				8.\hspace{4ex}\textbf{End For} (repeat within the same SI group)\\
			9.\hspace{2ex}\textbf{End For} (repetition for SI groups)\\
		\hspace{-1.1ex}10.~\textbf{End For}\\
		\hspace{-1.1ex}11.~\textbf{For} block $B=2:V$ \textbf{do}\\ 
		\hspace{-1.1ex}12.\hspace{2ex}\textbf{For} $DB_{B}=1:N$ \textbf{do} \vspace{-1ex} 
				\begin{equation*}
				\hspace{3.7ex} 
				\begin{aligned} 
		\hspace{-6.4ex}\text{13.} \quad\quad	Q^{[\nu]}_{B,R}(DB_{B},{\cal M})\leftarrow &\text{\bf{Exploit-SI}}\big(Q^{[\nu]}_{B-1,R}(DB_{B}\!+\!1,{\cal I}_{Nb})\\[-0.7ex]
				&\hspace{1.5ex}\cup\dots\!\cup Q^{[\nu]}_{B-1,R}(DB_{B}\!+\!Nb,{\cal I}_{1})\big) 
				\end{aligned} \end{equation*}
		
		\hspace{-1.1ex}14.\hspace{4ex}\textbf{For} side-information group $G=1:K$ \textbf{do} \\
		\hspace{-1.1ex}15.\hspace{6ex}\textbf{For} $RG=1:(R_{B}/K)$ \textbf{do} \vspace{-1ex}  
						\begin{equation*} \vspace{-0.6ex} 
				\hspace{-6.1ex}\text{16.} \qquad\qquad Q^{[\nu]}_{B,R}(DB_{B},{\cal I}_{G})\leftarrow \text{\bf{M-Sym}} (Q^{[\nu]}_{B,R}(DB_B,{\cal M}) )
						\end{equation*}
		\hspace{-1.1ex}17.\hspace{6ex}\textbf{End For} (repeat within the same SI group)\\
		\hspace{-1.1ex}18.\hspace{4ex}\textbf{End For} (repeat for SI groups)\\
			
		\hspace{-1.1ex}19.\hspace{2ex}~\textbf{End For} (repeat for each database)\\
		\hspace{-1.1ex}20.~\textbf{End for} (repeat for each block)\\
		\hspace{-1.1ex}21.~\textbf{For}  query round $R=2:K$ \textbf{do}\\
		\hspace{-1.1ex}22.\hspace{2ex}\textbf{For} $DB_1=1:N$ \textbf{do}\vspace{-1ex} 
		 \begin{equation*}\vspace{-0.5ex}\hspace{-5ex}
		 \begin{aligned}
		 \hspace{-6.2ex}\text{23.} \qquad\qquad& Q^{[\nu]}_{1,R}(DB_1,{\cal M})\leftarrow Q^{[\nu]}_{1,R-1}(DB_1-1,{\cal M})\\
		 \hspace{-6.2ex}\text{24.} \qquad\qquad& Q^{[\nu]}_{1,R}(DB_1,{\cal I}_G)\leftarrow Q^{[\nu]}_{1,R-1}(DB_1-1,{\cal I}_G)
		 \end{aligned}
		 \end{equation*}
		\hspace{-1.1ex}25.\hspace{2ex}\textbf{End For} (initializing rounds) \\
		\hspace{-1.1ex}26.\hspace{2ex}\textbf{For} block $B=2:V$ \textbf{do}\\
   			\hspace{-1.1ex}27.\hspace{3ex}\textbf{For} $DB_B=1:N$ \textbf{do}\\
   			\hspace{-1.1ex}28.\hspace{4ex}\textbf{For} side information group $G=1:K$ \textbf{do} \vspace{-1ex} 
   			  \begin{equation*} \vspace{-0.5ex} 
   			 	\begin{aligned}
   		\hspace{-7.0ex}\text{29.} \quad\qquad Q^{[\nu]}_{B,R}(&DB_B,{\cal I}_{G})\leftarrow Q^{[\nu]}_{B,R-1}(DB_B-1,{\cal I}_{G}) \\
   		\hspace{-7.0ex}\text{30.} \quad\qquad Q^{[\nu]}_{B,R}(&DB_B,{\cal M})\leftarrow \text{\bf{Reuse-SI}}\big(Q^{[\nu]}_{B,R}(DB_B,{\cal I}_{G}),\\[-0.7ex] 
   			 		&Q^{[\nu]}_{B-1,1}(DB_{B}+1,{\cal I}_{Nb+R-1})\cup\dots \\[-0.7ex]
   			 		&\quad\qquad\dots\cup Q^{[\nu]}_{B-1,1}(DB_{B}+Nb,{\cal I}_{R})\big)
   		 		\end{aligned} 
   		 	  \end{equation*}
   			\hspace{-1.1ex}31.\hspace{4ex}\textbf{End For} (SI groups) \\
			\hspace{-1.1ex}32.\hspace{3ex}\textbf{End For} (repeating for each database) \\
		\hspace{-1.1ex}33.\hspace{2ex}\textbf{End For} (repeating for each block) \\
		\hspace{-1.1ex}34.~\textbf{End For} (repeating for each round) \\
		\hspace{-1.1ex}35.~\textbf{For} $DB_B=1:N$ \textbf{do}\vspace{-1ex}
		 \begin{equation*}\hspace{1ex}\vspace{-0.5ex}
				\begin{aligned}
		\hspace{-2.1ex}\text{36.} \quad\quad\!	&Q^{[\nu]}(DB_B)\!\leftarrow\!\!\bigcup\limits_{B=1}^{V} \!\bigcup\limits_{R=1}^{K}\!\! \big(Q^{[\nu]}_{B,R}(DB_B,{\cal I})\cup Q^{[\nu]}_{B,R}(DB_{B},{\cal M}) \big)
				\end{aligned} \end{equation*}
		\hspace{-1.1ex}37.~\textbf{End For} (assembling the query sets)\\
		\specialrule{.1em}{.05em}{.05em}
	\end{tabular}
\normalsize
\vspace*{-4ex}
\end{table}

\vspace{-1ex}
\subsection{Sign assignment and redundancy elimination}\label{sec:SignAssyment} \vspace{-0.5ex}
We carefully assign an alternating sign $\sigma_t \in [+1,-1]$ to each symbol in the query set, based on the desired linear combination index $\nu$ \cite{SunJafararxiv_2017}. The intuition behind the sign assignment is to introduce a uniquely solvable linear equation system from the different $v$-sum types. By obtaining such an equation system in each block, the user can opt from downloading these queries, compute them off-line, and thus reduce the download rate. Based on this insight we can state the following lemma.

\begin{lemma}[\hspace{-0.1ex}\cite{SunJafararxiv_2017}]
	For all {$\nu\in[1:V]$}, each database $n\in[1:N]$, and based on the side information available from the neighboring databases, there are $V-M\choose v$ redundant $v$-sum types out of all possible types $V\choose v$ in each block {${v\in[1:V-M]}$} of the query sets.
\end{lemma}

Lemma~1 is also applicable when the desired linear function is performed over MDS-coded databases due to the fact that each MDS-coded symbol is itself a linear combination. That is, the MDS code can be seen as an inner code and the desired linear function as an outer ``code'' with respect to the databases. Hence, the redundancy resulting from the linear dependencies between messages is also present under MDS coding and we can extend Lemma~1 to our scheme. We now make the final modification to our PFR query sets. We first directly apply the sign assignment $\sigma_t$, then remove the redundant $v$-sum types from every block $B\in{[1:V]}$. Finally, we generate the query matrices $Q^{[\nu]}_{1:N}$ using a one-to-one mapping function $f,$ for which $Q^{[\nu]}(DB_B)$ is the preimage.

\begin{proof}
	The proof of optimality for arbitrary $N,K,M$ and $V$ follows from the structure of the query and Lemma~1. The achievable rate is given as 
	\begin{equation*}\label{eq:Cap_proof}
	\begin{split}
	&R\stackrel{(a)}{\leq}\frac{KN^V}{KN\sum_{v=1}^{V}\Big( {V\choose v}-{V-M\choose v}\Big) K^{V-v}(N-K)^{v-1}}\\
	&=\frac{N^V\big(\frac{N-K}{N}\big)}{\sum_{v=1}^{V}\Big( {V\choose v} K^{V-v}(N-K)^{v}-{V-M\choose v} K^{V-v}(N-K)^{v}\Big)}\\
	&\stackrel{(b)}{=}\frac{N^V\big(\frac{N-K}{N}\big)}{(N^V\!-K^V)-\sum_{v=1}^{V-M} {V-M\choose v} K^{V-v}(N-K)^{v}}\\
	&=\frac{N^V\big(\frac{N-K}{N}\big)}{(N^V\!-K^V)-K^M\sum_{v=1}^{V-M} {V-M\choose v} K^{V-M-v}(N-K)^{v}}\\
	&=\frac{N^V\big(1-\frac{K}{N}\big)}{(N^V\!-K^V)-K^M\big( N^{V-M}-K^{V-M}\big)}\\
	&=\frac{N^V\big(1-\frac{K}{N}\big)}{(N^V\!-K^V)-K^M N^{V-M}+K^{V}}\\
	&=\frac{N^V\big(1-\frac{K}{N}\big)}{N^V\!-K^M N^{V-M}}=\frac{1-R_c}{1-R_c^M};
	\end{split}
	\end{equation*}
	where  (a) follows from the definition of the PFR rate
        \eqref{eq:PFR_rate_def}; (b) follows from the fact that the second
        term of the summation in the denominator is equal to zero for
        $v>V-M$ and consequently we can change the upper bound of the summation; and the first term of the summation follows from the binomial theorem.
\end{proof}

\vspace{-1.5ex}
\subsection{Correctness (decodability)}\vspace{-0.5ex}
To prove correctness, we show that the user can obtain the desired linear combination $\widetilde{W}_\nu$ from the answers retrieved from $N$ databases. From the query answers $A^{[\nu]}_{1:N}$, we group the $K$ identical queries from different rounds and databases. Each group will result in $K$ linearly independent equations that can be uniquely solved. We decode, block by block, starting from block one, which we directly decode and obtain 
$K N \big(\!{V\choose v}-{V-M\choose v}\!\big) K^{V-v}(N-K)^{v-1}$ decoded symbols. Now, using these symbols we regenerate ${V-M\choose v}$ redundant symbols according to Lemma~1 and obtain 
{$K N {V\choose v} K^{V-v}(N-K)^{v-1}$} symbols in total. Out of these queries there are 
$K N \big(\!{V\choose v}-{V-1\choose v}\!\big) K^{V-v}(N-K)^{v-1}$ symbols from $\widetilde{W}_\nu$. 

Next, for blocks $B\in[2:V]$, we use the symbols obtained in the previous block $B-1$ to remove the side information associated with the desired linear combination symbols of the current block $B$, then the operations from the first block (decode and retrieve redundancy) are repeated. As a result, we obtain a total number of symbols equal to $K N\sum_{v=1}^{V}\big(\! {V\choose v}-{V-1\choose v}\!\big) K^{V-v}(N-K)^{v-1} =\frac{N}{N-K} \big( N^V- KN^{V-1}\big) = K N^V$ denoting precisely the number of symbols in $\widetilde{W}_\nu$.

\vspace{-0.5ex}
\subsection{Privacy}
\vspace*{-0.5ex}
Privacy is guaranteed by preserving an equal number of requests for any
linear combination $\widetilde{W}_\nu,$ where the requests are symmetric from the
perspective of the accessed virtual messages. As the MDS code can be seen as
an outer code, the arguments in \cite{Maddah-Aliarxiv2017,
  SunJafararxiv_2017} apply here as well. In particular, each database is queried with precisely the same $v$-sum type
components, i.e., $U_\nu(t)$, which ensures symmetry. This can be seen from Step 16 in Algorithm 1 where the same subroutine ``M-Sym'' is used for each block and database. By selecting  a permutation $\pi(t)$ and a sign assignment $\sigma_t$ uniformly at random, queries for code symbols are permuted in the same way over all databases. With other words, for any
$U_\nu(t)=\sigma_t {\bf v}_\nu {\bf W}[\pi(t)]$ there exist $\sigma_t,
\pi(t)$ such that  $Q^{[\theta]}(DB_B) \leftrightarrow Q^{[\nu]}(DB_B) \quad
\forall \nu, \theta \in [1:V]$. Thus, $A_n^{[\nu]}$, and $Q_n^{[\nu]}$
are statistically independent of $\nu$ and \eqref{privacy_const} holds.


\vspace{-1ex}
\section{Example} \label{EX} \vspace{-0.5ex}
We consider $M=2$ messages stored using a $(3,2)$ MDS code. The user wishes to obtain a linear combination over the binary field (i.e.,~${\bf{v}_\nu}\in {\mathbb F}_2^M, \quad V=3$). Therefore, we have the linear combinations ${{\bf{v}}_1=[1 \;0], {\bf{v}}_2=[0 \; 1], {\bf{v}}_3=[1\; 1]},$ and each message must be of length $L=KN^V=54$ symbols. We simplify the notation by letting $a_t=U_1(t), \; b_t=U_2(t),$ and $c_t=U_3(t)$ for all $t\in[1:27]$. Let $\sigma_t=1\;\forall t$ and let the desired linear combination index be $\nu=3$. 

\indent{\textit{Query set construction:}}
Algorithm~1 starts with $B=1$ by generating queries for each database and {$K^{V-1}=4$} distinct instances of $c_t$ (i.e.,~from database~1 query ${\bf{g}}_1^T(c_{1:4})\triangleq\{{\bf{g}}_1^Tc_1,\dots, {\bf{g}}_1^Tc_4\}$). By message symmetry this also applies for $a_t$ and $b_t$ to form two groups of side information sets to be used in the next block with $NK^{V-1}=12$ symbols in total from each linear combination. Next, one group of side information is queried jointly with a new instant of the desired message. For example, for database~1 and type $b+c$ we have ${\bf{g}}_1^T(b_{5:6}-c_{13:14}) \triangleq  \{{\bf{g}}_1^Tb_5-{\bf{g}}_1^Tc_{13}, {\bf{g}}_1^Tb_6-{\bf{g}}_1^Tc_{14}\}$. The remaining blocks and rounds follow from Algorithm~1. After generating the query set for each database, we apply the sign assignment and remove the redundant queries. 

\indent{\textit{Decoding:}}
The answer strings from each query are shown in Table~1. Note that there is
no $c_t$ in the first block as they are redundant and can be generated by
the user. To decode we start with Block 1, and we obtain ${\bf{g}}_1^Ta_1$
from database~1 and ${\bf{g}}_2^Ta_1$ from database 2. Thus, by the MDS code
properties we can decode and obtain the segment $a_1$; similarly for all
other queries in this block. Now for Block~2, we first remove the side
information from the types containing symbols of $c_t$. For example, for
${\bf{g}}_1^Tb_5-{\bf{g}}_1^Tc_{13}$ from database~1, we have $b_5$ from the
previous block. As a result, we obtain ${\bf{g}}_1^Tc_{13}$. Similarly we
obtain ${\bf{g}}_2^Tc_{13}$ from database~2, and $c_{13}$ can be recovered.

\renewcommand{\baselinestretch}{0.5}
\begin{table}[htbp]
\centering
\begin{minipage}{0.5\textwidth}
{\linespread{1} 
\vspace{-0.8ex} 
\caption{ The query response for PFR from $(3,2)$ MDS coded databases, $M=2$, $V=3$, and $\nu=3$}}
\vspace{-2.2ex}
\vskip 0.03125in
\label{table:QR}
\footnotesize 
\begin{tabular}
{|@{\,}m{5ex}@{\;}|@{}c@{\;}|@{\;}c@{\;}|@{\;}c@{\;}|} 
\hline
\rule{0pt}{1pt}& & & \vspace*{-0.5ex}\\
{\scriptsize($R$,$B$)}& {DB1} & {DB2} & {DB3}\vspace*{-0.5ex}\\	
\rule{0pt}{1pt}& & & \\ 
\hline	
\rule{0pt}{1pt}& & & \vspace*{-0.5ex}\\	
\multirow{2}{*}
{\scriptsize{$(1,1)$}}&
{${\bf{g}}_1^T(a_{1:4})\!$}&
{${\bf{g}}_2^T(a_{5:8})\!$}&
{${\bf{g}}_3^T(a_{9:12})\!$}\\
&
{${\bf{g}}_1^T(b_{1:4})$}&
{${\bf{g}}_2^T(b_{5:8})$}&
{${\bf{g}}_3^T(b_{9:12})$}\vspace*{-0.5ex}\\
\rule{0pt}{0pt}& & & \\ 
\hline
\rule{0pt}{1pt}& & & \vspace*{-0.5ex}\\
\multirow{3}{*}
{\scriptsize{$(1,2)$}}
\multirow{2}{*}
&
{${\bf{g}}_1^T(b_{5:6}-c_{13:14})$}&
{${\bf{g}}_2^T(b_{9:10}-c_{17:18})$}&
{${\bf{g}}_3^T(b_{1:2}-c_{21:22})$}\\
&
${\bf{g}}_1^T(a_{5:6}-c_{15:16})$&
${\bf{g}}_2^T(a_{9:10}-c_{19:20})$&
${\bf{g}}_3^T(a_{1:2}-c_{23:24})$\vspace*{-0.5ex}\\
\rule{0pt}{1pt}& & &\\ 
\cline{2-4}
\rule{0pt}{1pt}& & &\vspace*{-0.5ex}\\
&
$\;{\bf{g}}_1^T(a_{13:14}-b_{15:16})$&
$\;{\bf{g}}_2^T(a_{17:18}-b_{19:20})$&
$\;{\bf{g}}_3^T(a_{21:22}-b_{23:24})$\vspace*{-0.5ex}\\
\rule{0pt}{1pt}& & &\\ 
\hline 
\rule{0pt}{1pt}& & &\vspace*{-0.5ex}\\
{\scriptsize{$(1,3)$}}&
${\bf{g}}_1^T(a_{17}\!-\!b_{19}\!+\!c_{25})$&
${\bf{g}}_2^T(a_{21}\!-\!b_{23}\!+\!c_{26})$&
${\bf{g}}_3^T(a_{13}\!-\!b_{15}\!+\!c_{27})$\vspace*{-0.5ex}\\
\rule{0pt}{1pt}& & &\\ 
\hline
\rule{0pt}{1pt}& & &\vspace*{-0.5ex}\\
\multirow{2}{*}
{\scriptsize{$(2,1)$}}&
{${\bf{g}}_1^T(a_{9:12})$}&
{${\bf{g}}_2^T(a_{1:4})$}&
{${\bf{g}}_3^T(a_{5:8})$}\\
& 
{${\bf{g}}_1^T(b_{9:12})$}&
{${\bf{g}}_2^T(b_{1:4})$}&
{${\bf{g}}_3^T(b_{5:8})$}\vspace*{-0.5ex}\\
\rule{0pt}{1pt}& & &\\ 
\hline
\rule{0pt}{1pt}& & &\vspace*{-0.5ex}\\
\multirow{3}{*}
{\scriptsize{$(2,2)$}}
\multirow{2}{*}
&
{${\bf{g}}_1^T(b_{7:8}-c_{21:22})$}&
{${\bf{g}}_2^T(b_{11:12}-c_{13:14})$}&
{${\bf{g}}_3^T(b_{3:4}-c_{17:18})$}\\
&
{${\bf{g}}_1^T(a_{7:8}-c_{23:24})$}&
{${\bf{g}}_2^T(a_{11:12}-c_{15:16})$}&
{${\bf{g}}_3^T(a_{3:4}-c_{19:20})$}\vspace*{-0.5ex}\\
\rule{0pt}{1pt}& & &\\ 
\cline{2-4}
\rule{0pt}{1pt}& & & \vspace*{-0.5ex}\\
\multirow{2}{*}
&
$\;{\bf{g}}_1^T(a_{21:22}-b_{23:24})$&
$\;{\bf{g}}_2^T(a_{13:14}-b_{15:16})$&
$\;{\bf{g}}_3^T(a_{17:18}-b_{19:20})$\vspace*{-0.5ex}\\
\rule{0pt}{1pt}& & &\\ 
\hline 
\rule{0pt}{1pt}& & &\vspace*{-0.5ex}\\
{{$(2,3)$}}&
${\bf{g}}_1^T(a_{18}\!-\!b_{20}\!+\!c_{27})$&
${\bf{g}}_2^T(a_{22}\!-\!b_{24}\!+\!c_{25})$&
${\bf{g}}_3^T(a_{14}\!-\!b_{16}\!+\!c_{26})$\vspace*{-0.5ex}\\
\rule{0pt}{1pt}& & &\\ 
\hline
\end{tabular}
\end{minipage}
\end{table}
\renewcommand{\baselinestretch}{1}
\vspace{-2ex}
\section{Outer Bound for the  Special Case $V=2$}
\vspace{-0.5ex}
In the special case of $V=2,$ $M$ independent messages, and any $(N,K)$ MDS
code an outer bound for the coded PFR problem is obtained by combining the
independence of answer strings from any $K$ databases \cite[Lemma
2]{BanawanUlukus_Globecom2017} with 
\cite{SunJafararxiv_2017}. Thus, we can show that the retrieval rate is
upper 
bounded as $R \leq
\frac{NH(\omega_1)}{KH(\omega_1,\omega_2)+ H(\omega_1)(N-K)}$, where the
joint distribution of $(\omega_1,\omega_2)$ is the joint distribution of
$(\widetilde{W}_{1,\ell}, \widetilde{W}_{2,\ell})$ for all $\ell \in [1:\tilde{L}]$ selected iid~with respect to the symbols of the messages.

\vspace{-1ex}

\bibliographystyle{IEEEtran}
\bibliography{IEEEabrv,references}
\end{document}